# Lights and shadows of COVID-19, Technology and Industry 4.0

**Nicola Melluso**
Department of Energy Systems, Territory and Construction Engineering
University of Pisa
Pisa, 56126
nicola.melluso@phd.unipi.it

**Silvia Fareri**
Giacomo Brodolini Foundation, Marco Biagi Foundation
University of Modena e Reggio Emilia

**Gualtiero Fantoni**
Department of Civil and Industrial Engineering
University of Pisa

**Andrea Bonaccorsi**
DESTEC
University of Pisa

**Filippo Chiarello**
DESTEC
University of Pisa

**Pietro Manfredi**
Department of Civil and Industrial Engineering
University of Pisa

**Elena Coli**
Department of Information Engineering
University of Pisa

**Vito Giordano**
Department of Information Engineering
University of Pisa

**Shahin Manafi**
Giacomo Brodolini Foundation, Marco Biagi Foundation
University of Modena e Reggio Emilia

## Abstract

Scientific discoveries and technologies played a significant role in the digital revolution that occurred over the last years. But what is their role in the turmoil brought by the current pandemic? The aim of this paper is to show how digital technologies are operating during this first phase of the spreading of COVID-19. The study analyses and debates the current and potential role of digital technologies, focusing on their influence in the industrial and social fields. More specifically we used the blogging platform "Medium", which has seen an exponential growth in its production of articles over the last couple of months. Even if different from esteemed scientific sources, this platform provides a structure that suits our analysis. We searched how many times digital technologies are mentioned in articles regarding Coronavirus and, after collecting these articles, we collected page tags (comparable to "keywords" in scientific articles) and classified them (technology tags and non-technology tags) to create a graph showing the relation between them. This network allowed us to acknowledge and picture how technologies are currently debated. This was the starting point to discuss the key implications for an imminent future, and question about the impact on industry, society and labour market. What are the opportunities or threats of using technologies of Industry 4.0? Which are the needs rising because of the pandemic and how can technologies help in their fulfillment? How will the industrial scenario change after this pandemic? How will the labour market be affected? How can technologies be advantageous in the emerging social challenges?

# 1 Introduction

We are living a world in which digital technology is playing a main role in development, growth and evolution, in particular for companies and activities involved in the process of producing goods for sale. Before the advent of this pandemic, we were witnesses of a gradual spreading and embracing of technologies connected to Industry 4.0 phenomenon. However, we can observe how a change is happening in the current situation and foresee how differently technologies will be employed in the future. Both industry 4.0 and the pandemic present themselves as two revolutionary phenomena that lead to radical changes.

The aim of this study is to discuss the role of digital and 4.0 technologies in relation to their impact on industry, labour market and society during Covid 19 pandemic. We use as starting point for this discussion a quantitative analysis of how these technologies are behaving in this pandemic scenario. In particular, we looked at how much 4.0 technologies are mentioned in the articles about coronavirus in Medium's blogging platform. By linking the technologies found with the topics discussed in the articles, it is possible to create a data driven relational structure that allows us to carry out our discussions.

At this stage of the pandemic, the efforts of scientific production are mainly focused on finding a cure and how to manage the emergency situation. However, we cannot overlook what will come after the end of this extraordinary situation. Therefore, the collection of information must take into account a broad and in-time knowledge source as Medium, in order to be able to frame the scenario perfectly. There are numerous sources of information and numerical data on how the virus is spreading and impacting industries and the economy in general, for instance papers. However, given the fast pace of the diffusion and the recent nature of the virus we cannot rely on documents that have long-waiting periods before approval since our objective is to analyse data that is already available and thus analysable.

This study is a first attempt to demonstrate how the advent of COVID-19 did not stop the advance of technologies 4.0, in fact it gave it a push, a drive that certainly has a change in its direction. If some technologies are maybe becoming inadequate and inefficient in this emergency scenario, some have seen an increase in their employment and a massive development.

The paper is organized as follows. Section 2 gives an overview of the background of this study. In particular Section 2.1 discusses the Industry 4.0 phenomenon focusing on the technologies that played an important role in this paradigm shift; Section 2.2 shows how the Medium platform works and it explains in detail the reason we choose it as source of data. Section 3 presents the methodology used to gather, analyse and visualize data to capture relations between technologies and COVID-19. In Section 4 the results and outcomes of the methods are shown and described. Finally, Section 5 discusses how the role and impact of technology has changed in the face of this crisis on industry, the labour market and society.

# 2 Background

In this section we show past works that support our methodological approach. In particular, section 2.1 shows recent trends in industry 4.0 and efforts in the mapping and understanding of this fuzzy field; section 2.2 explores how Medium has previously been used in literature as a data source.

## 2.1 Industry 4.0 and Technologies

In the last few years the production system in advanced countries has been theater of a new paradigm shift, defined as Industry 4.0. The term Industry 4.0 has been initially formulated in Germany in 2011 by two researchers during the Hannover Trade Fair, to connote the current phase of computerization of the manufacturing systems [1]. It refers to a novel approach to the industrial system, that is based on the real-time connection of people, machines and objects for the intelligent management of logistic-production systems [2]. The opportunities and challenges resulting from this paradigm are such that today we are talking about the fourth industrial revolution.

It is said to be the fourth because this revolution was preceded by three other revolutions: the first dates back to the early nineteenth century with the invention of the steam engine and the exploitation of its power for the mechanization of production; the second (end of the 19th century) saw the spread of the assembly line and the concept of dividing the work aimed at mass production; the third focused on the introduction of automated systems based on electronic and information technologies [3]. The growing interest in Industry 4.0 in everyday life is confirmed by the increasing number of articles focusing on topics that are related to the “Industry 4.0”. An example of this trend is shown in Figure 1 which displays the number of times that “Industry 4.0” has been used as a tag in Medium articles, i.e. 2026 times in the last 4 years. The growth has been approximated by the line.

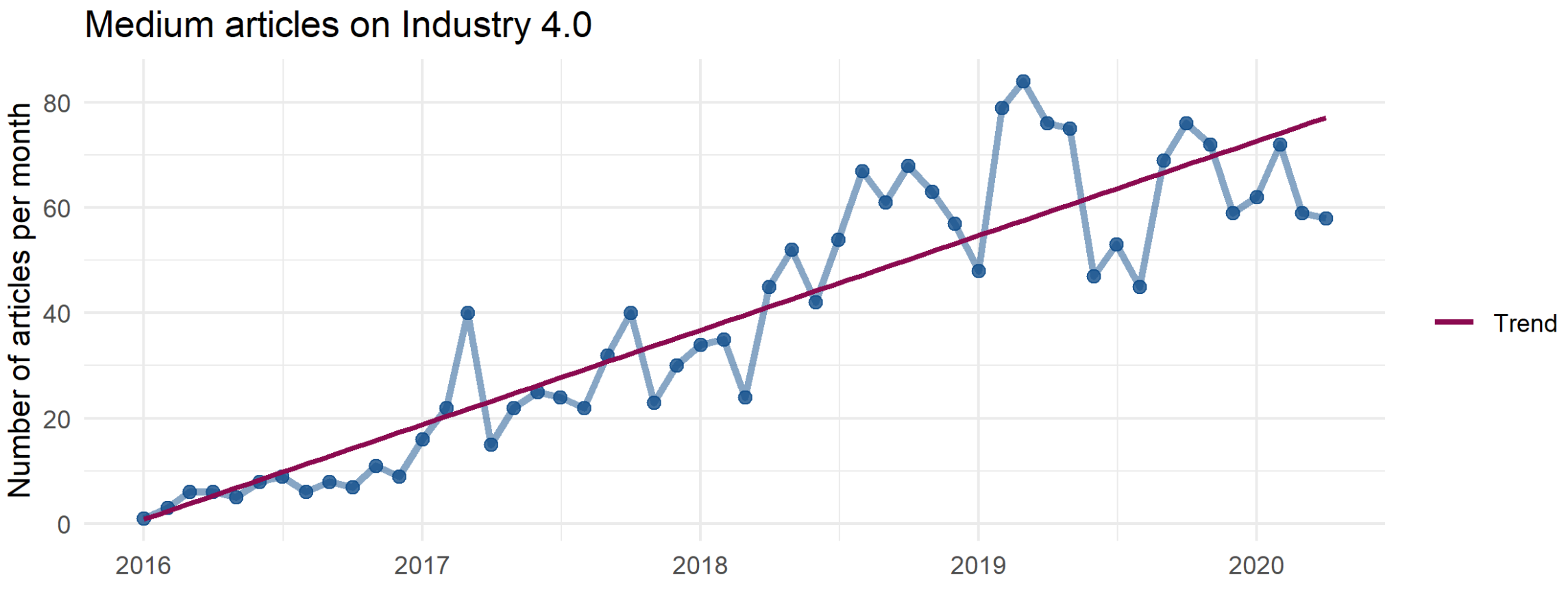

Figure 1: Number of Medium articles with the tag “Industry 4.0”

The role of the technologies has been considered central in the current landscape of Industry 4.0, in fact many authors speak about industry 4.0 as a "tech tsunami". In particular, we can see the fourth industrial revolution as an intelligent recombination of existing technologies and some new technologies and their application in the manufacturing environment [4]. This definition could be misleading, suggesting that the wave influences only manufacturing companies, however, Industry 4.0 affects all sectors, also the Public Administration and the everyday life of citizens.

As said above, the technologies are the key elements of the 4.0 paradigm, therefore a question arises: “What are the main technologies of the Industry 4.0 paradigm?”. While there is a reasonable convergence on the definition of Industry 4.0, there is still disagreement and misalignment with respect to the 4.0 technologies [5], caused by the large amount of possible applications that the new paradigm affects [6]. This issue has been faced with two concurrent top-down and bottom-up approaches.

The top-down approach involved a panel of experts employed to map and define the Industry 4.0 technological landscape from different points of view (e.g. academic or industrial perspective). An example of these top-down studies is the classification of technologies of McKinsey, Boston Consulting Group or Xu, L. (2018). They classify the technologies in various categories. For example the Boston Consulting Group clusters the 4.0 technologies into 9 categories: autonomous robots; simulation; horizontal and vertical system integrator; internet of things; cybersecurity; cloud; additive manufacturing; augmented reality; big data and analytics.

On the other hand, the bottom-up approach used text mining techniques to extract the technologies from a corpus of text. In this context, it is interesting the work of Chiarello, F. et al. (2018) that leverages publications and open source repositories (Wikipedia) to define and cluster the technologies of the Industry 4.0 paradigm. The researchers were able to find more than 1200 technologies of the Fourth Industrial Revolution and they have grouped them into 11 clusters: big data; transactions, digital certification, digital currency; programming languages; computing; embedded systems; intel technologies; internet of things; protocols and architectures; communication networks and infrastructures;production technologies; identification technologies. Both top-down and bottom-up approaches demonstrated that Industry 4.0 could be mapped in a technological domain.

Governments and policy departments define their guidelines referring to the Key Enabling Technologies (KETs). An example is given by the Framework Programmes for Research and Technological Development (H2020) that defines KETS as “the driver for economic growth and employment to improve the EU’s competitiveness in a rapidly changing global environment”.

### 2.2 Data Source: Medium

At the state of the art, the largest free and open repository of knowledge for COVID-19 is CORD-19 [1] implemented by Allen Institute for AI and containing over 52.000 scholarly articles, including around 41.000 full text, about the coronavirus family of viruses and, in particular, about COVID-19. Although CORD-19 is the most widely used repository for scientific production about COVID-19, the focus of most articles is on the medical perspective. The result is that information about technologies linked to COVID-19 are almost impossible to be found within the repository.

[1] https://pages.semanticscholar.org/coronavirus-research

Therefore, the idea is that, in this specific context and for our specific purpose, scientific literature is not the best source to be used. So, we focused on not-scientific sources and, in particular, Medium.

Medium is one of the most trending blogging web-platforms. Statistics show almost 60 million monthly readers at the end of 2017, 7,5 million posts and 2 billion words written on Medium in 2016[2]. The actual estimated numbers are even higher. The main difference between Medium and the other blogging platforms is that Medium is more social-oriented. In fact, it allows some following and liking features that make it really close to a social network. Therefore, Medium could be considered as a hybrid between scientific articles and Twitter, as explained below. Every Medium article, before being published, should be associated with one or more tags in a similar fashion to keywords for an academic paper.

The use of Medium brings different advantages and disadvantages. Among the first, we could mention the availability of a large number of full texts. Moreover, the availability of keywords associated with Medium's articles is crucial for further analysis carried out in this work. For what concerns Medium's disadvantages, they are mainly linked to the reliability of the articles, since their review process is very fast and not structured.

The chart in Figure 2 shows how much the topic of Covid-19 is spreading on Medium. Confronting Figure 1 and Figure 2, it is visible how Medium has answered fast to the Covid-19 pandemic, reaching a level of interest far behind Industry 4.0 already before March 2020 when also European countries started the lock-down.

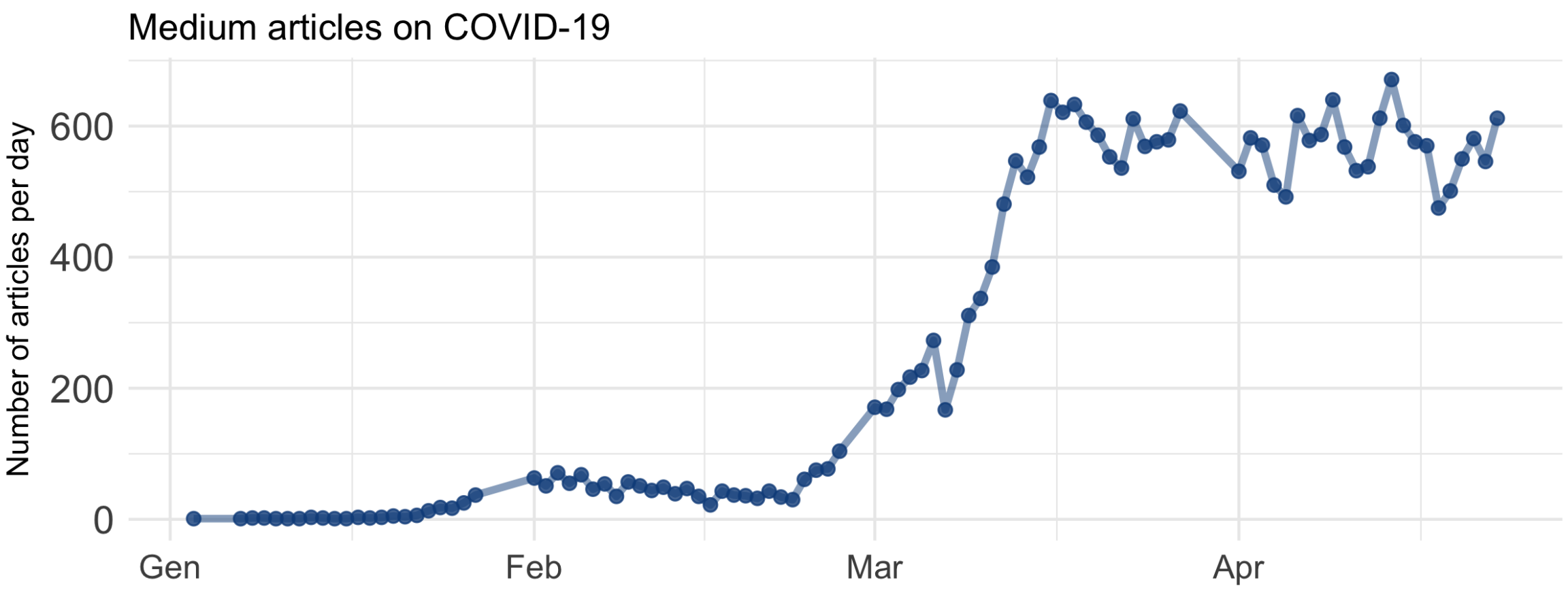

Figure 2: COVID-19 topic spread on Medium

## 3 Methodology

As said in the introductory section, the scope of the present study is to analyse the connections between technologies and the COVID-19 and how the latter is influencing these links. Given this purpose, the adopted methodology is structured as follows:

- definition of the appropriate data source to extract the necessary information;
- retrieval of all documents related to COVID-19;
- extraction of digital technologies from the documents;
- analysis of how technologies are connected to the other topics discussed within the documents.

Each phase is explained in the following paragraphs.

**Definition of the data source** There are numerous sources of information and numerical data on how the virus is spreading and impacting industries and the economy in general, for instance papers. However, given the fast pace of the diffusion and the recent nature of the virus we cannot rely on documents that have long-waiting periods before approval since our objective is to analyse data that is already available and thus analysable. That being said, we found that Medium highly satisfies our requirements given that users have the possibility to publish articles without necessarily

[2]source: DMR (https://expandedramblings.com)

waiting for approvals and reviews. Moreover, like papers published on well known databases, e.g. Scopus, every article on Medium has specific tags which make it easier to understand the topics that are being addressed.

**Document Retrieval** After having identified the right data source for the study, we proceeded with the retrieval of all documents inherent to the pandemic. More specifically, we searched for articles that had the *covid-19* tag and subsequently, used web scraping techniques to download the related content, i.e. title of the articles, body and tags. This process was mainly facilitated by the structure of Medium's web pages.

**Extraction of technologies** Document retrieval is then followed by the extraction phase in which we adopted text mining algorithms to search for technologies belonging to a predefined list within the content of the downloaded articles. We use the complete list of technologies extracted and mapped by Chiarello et a. 2018. In particular we used regular expressions in order to extract also inflections (e.g. singular, plurals) of the technologies. This allowed us to have an initial understanding of which technologies are currently being mentioned together with covid-related articles, however it still does not provide a complete overview on their connections with the other topics.

**Analysis of connections** Finally, after extracting the technologies it was possible to define their link with the other topics addressed within the same articles. In detail, this was achieved by calculating the co-occurrence degree (i.e. how many times two elements appear together in the same document) between technologies and those tags which are not considered technologies. The counts were then represented as an (N, N) adjacency matrix, where N is the number of unique topics/technologies and the elements in the matrix indicate the number of co-occurrences. We then used the adjacency matrices to generate an undirected graph *Gs = (Vs, Es)*, where the vertices, *Vs*, are the topics/technologies, and the edges, *Es*, are the respective co-occurrences. This first graph has been visualized using Gephi [7] and it is shown in the Results section in Fig. 5.

Furthermore, we decided to take the analysis one step further by using web scraping to extract those technologies that Medium considers to be related to every technology of the graph. We then re-calculated the co-occurrence of the technologies and used it to create a new graph shown in Fig. 6. The latter, together with the one in Fig. 5, can provide valuable insight into technologies that might appear in articles in the near future regarding the COVID-19 based on the connections that technologies of the graph in Fig. 5 have with other technologies shown in Fig. 6. These results will be better analysed in the following section.

## 4 Results

The analysis has been done by extracting COVID-19 related articles for a time window ranging from 1st of January 2020 till the 17th of April 2020. We found a total number of 26.788 articles related to COVID-19, of which 453 containing at least one digital technology contained in the list. As we expected, the number of articles is having an exponential growth curve.

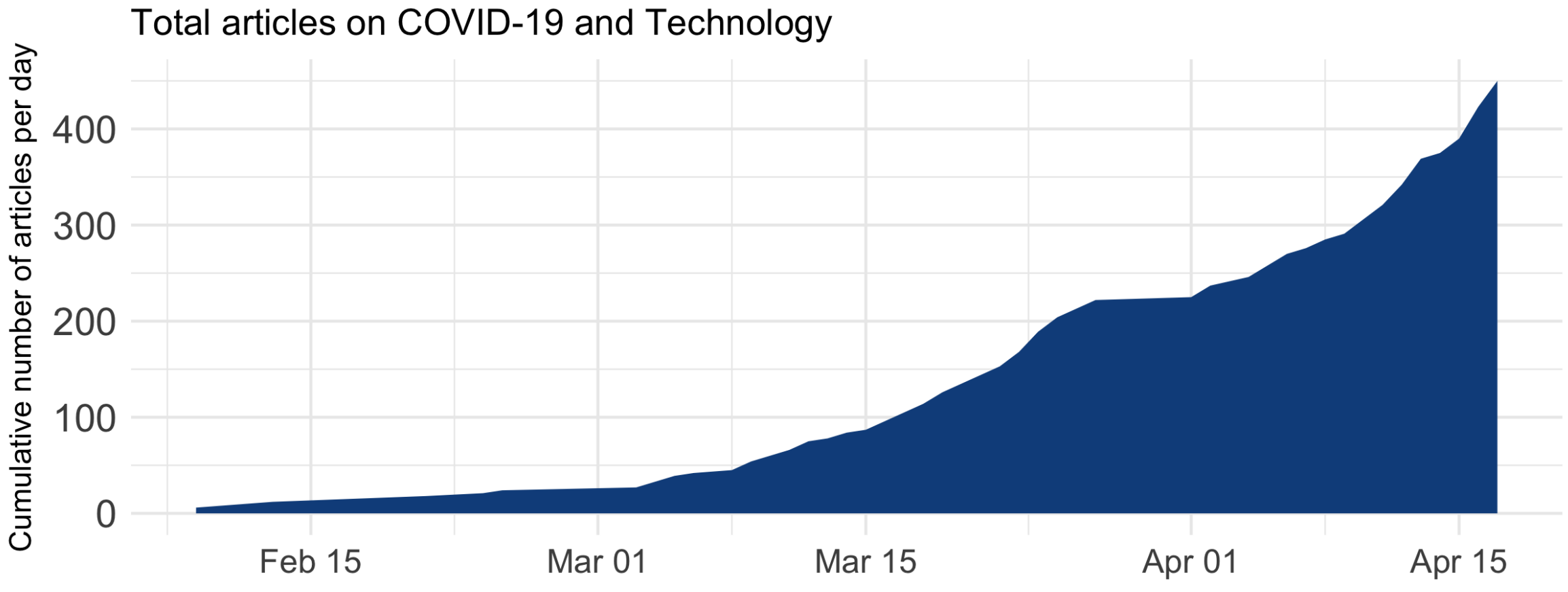

Figure 3: Number of articles of Tech-COVID

Figure 4a and 4b show a first view of the relationship between tags that are technologies ("Tech Tags") and tags that are not technologies ("Other Tags"). The values in the heatmap of Figure 4a represent the co-occurrence of two tags, namely the number of times are used in the same articles. Clearly, the “Surveillance” tag is the one that is highly cited with all the technologies. This is due to the fact that the use of geolocalization systems for surveillance has been subjected to an intense debate. It is also interesting to underline the high correlation between the tag “Social Distancing” and “Internet of things”. This is probably due to the fact that the use of this technology will face the emerging need of maintaining the social distance.

Figure 4b shows the co-occurrence values normalized by the variable “Other Tag”. This picture has to be read by looking at the values for each column. For example, the high value between “Cryptocurrency” and “Finance” means that for the “Finance” tag, the most cited technology is “Cryptocurrency”.

Further considerations could be done looking at the Figure 4b where the co-occurrence value is normalized using the minmax normalization by grouping the tags:

- the “Citizen” macro-cluster (“Social Distancing”, “Self-improvement”, “Mental Health”, “Public Health”) is frequently linked to “Internet of Things” and “Predictive Analytics”; this confirms that IoT is giving the main technological support to the citizens;
- the “State” macro-cluster (“Society”, “Politics”, “Economy”) shows both a focus on geolocalization technologies (and therefore to ensure the tracking of movements), and on technologies to support communication and effective information sharing;
- the “Business” macro-cluster (“Remote Working”, “Start-up”, “Innovation”), could be seen from the point of view of both worker and company; this cluster relies on the cloud as a decisive support, as well as remote control and virtual reality; it also interesting the high correlation with cybersecurity, probably due to the increasing amount data whose sharing is spreading.

Figure 5 shows the graph of the tags relations. The blue nodes represent the technologies, while those in green are the tags that are not technologies. The size of nodes is proportional to the number of times the tag is used in COVID-19 related articles. The layout of the graph (relative position of the nodes) is computed considering the number of times two tags are cited simultaneously in different articles: the more they are cited together, the closer two appear. For example, "quarantine" and "lockdown" are close because it is common to use the two tags for the same article. In this picture we removed the obvious tags of "pandemic", "covid" in order to have a clearer visualization of the results.

Figure 6 shows the graph of tags generated starting from the technologies found in the COVID-19 articles. Similar to the graph in Figure 5, each node is a tag and the size of the node depends on the in and out degree of each node. In practice, the larger is the node, the higher is related to the other tags. The main difference in this version of the graph, with respect to the one showed in Figure 5, is that the number of COVID articles in which these technologies are contained is negligible. In other words, the graph gives a picture of the technologies prior to COVID-19 manifestation. The image, if compared with the previous one (Figure 5), gives us a picture of the reconversions of technologies in COVID perspective.

An interesting example is given by the tag "Virtual Assistant". In Figure 6 it is linked to the term "Productivity" while in Figure 5 it is very close to the concept of health. This means that due to the pandemic, the technology could find more health related applications. Similarly, the behaviour of the tags of "Remote Control" and "Condition Monitoring" is crucial to understand the role of Industry 4.0. These two concepts appear highly in both graphs (of Figures 5 and 6) and their relation is given by the tag of “Smart Manufacturing” or “Internet of Things”.

## 5 Discussion

In the present session we discuss the results shown in section 4. We structured this section focusing on the joint impact that Digital Technologies and COVID-19 are having on Industry, labour market and Society as a whole respectively in sub-sections 5.1, 5.2 and 5.3.

### 5.1 COVID, Technologies and their impact on Industry

The post COVID-19 era will bring us into the “new normal” where temperature scanning, social distance, sanification, mask wearing and tele-whatever will be the way of living expecting us for the next 18-24 months. If some technologies were adopted only in airports, embassies, and few other sensitive places, now most of them can be adopted and many barriers to entry are falling down.

The use of teleconferences in judgments, the advent of *’ePrescription’* instead of medicinal prescription on paper sheets have been adopted in a few days in many countries around the world thus cancelling long debates and ancient

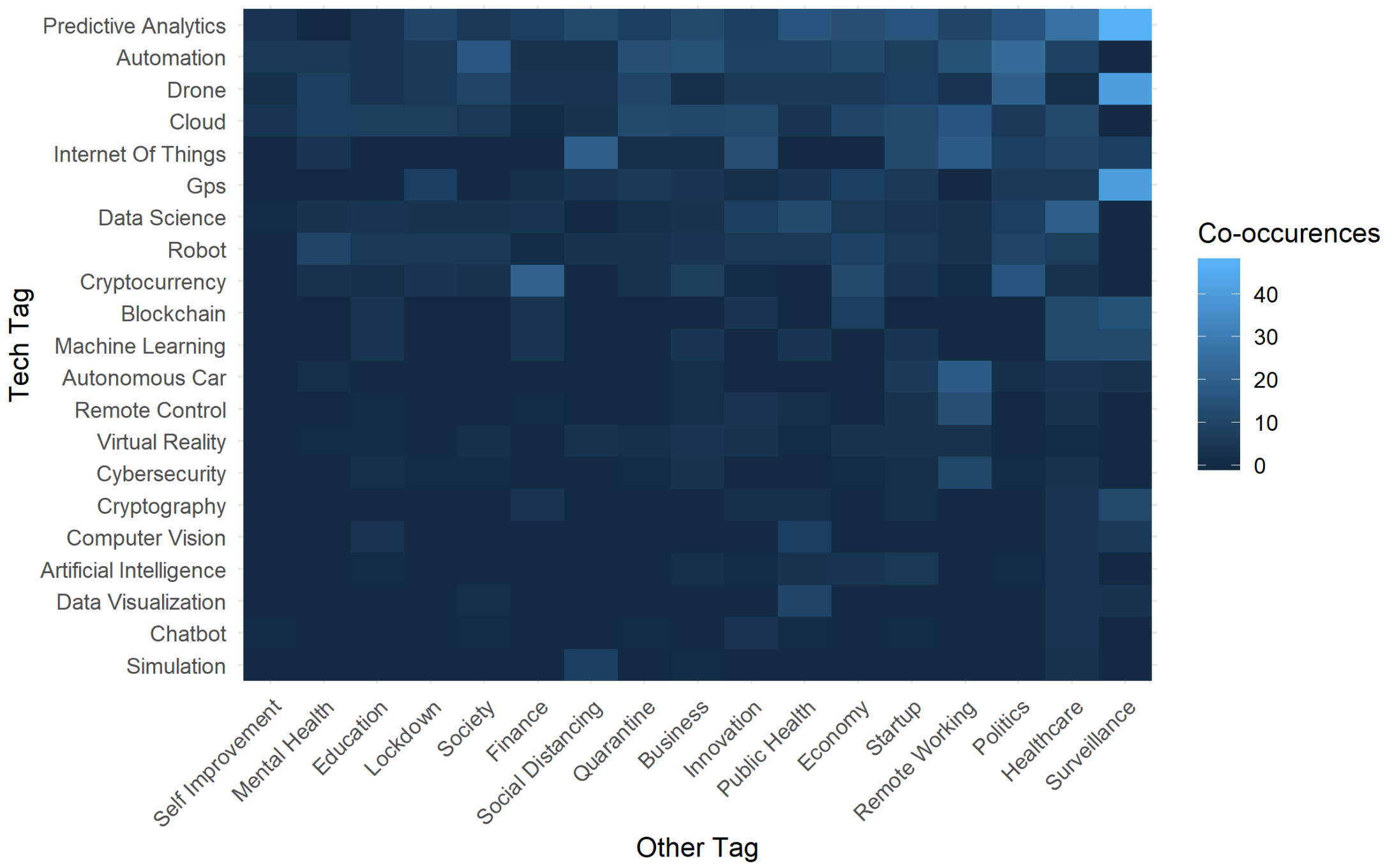

(a) Heatmap of technologies associated with COVID related tags

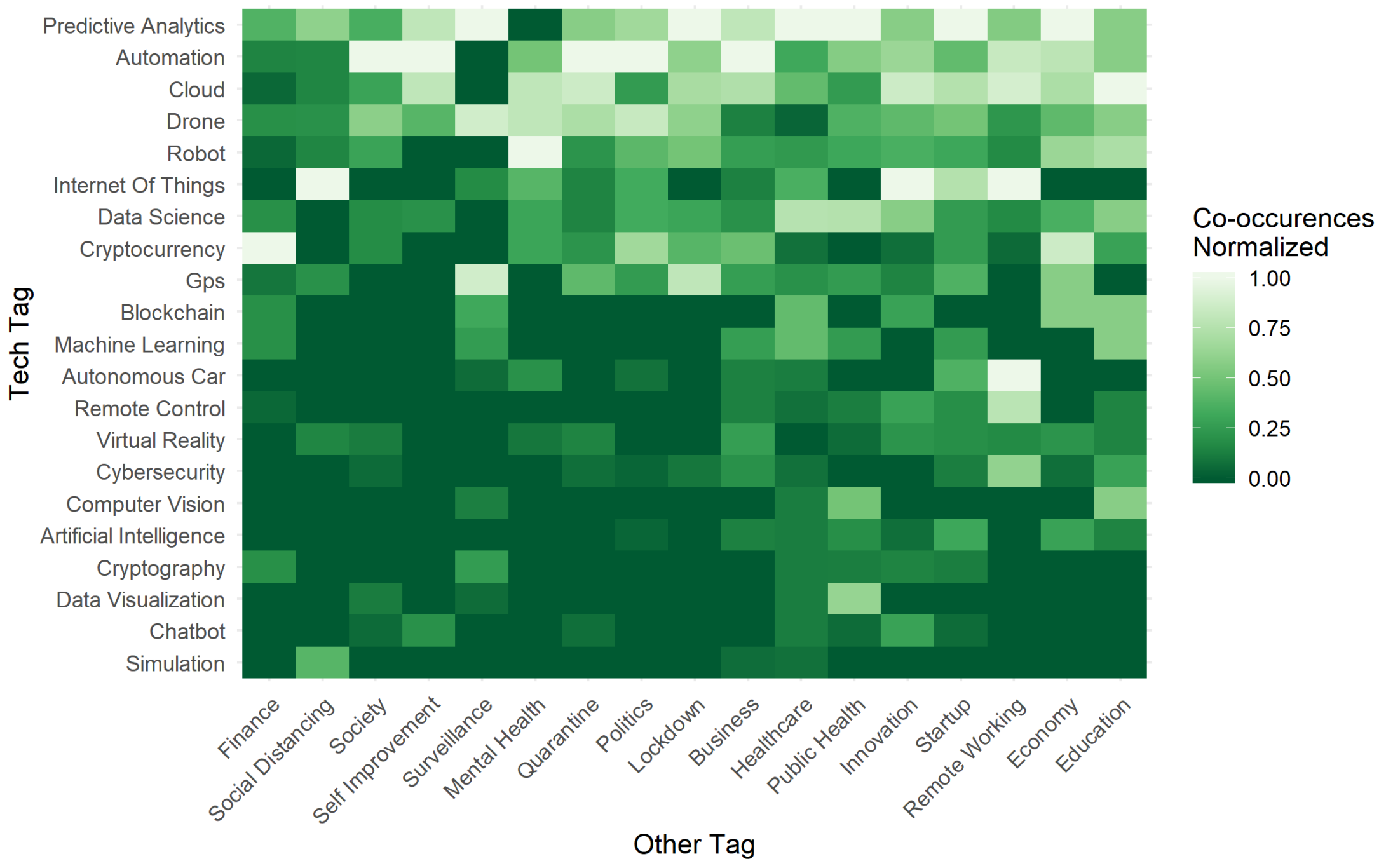

(b) Normalized heatmap of technologies associated with COVID related tags. The value is normalized using the minmax normalization by grouping the tags (Other Tag variable) resistances. Probably the same will happen with the apps for contact tracing in everyday life where the standard aversion towards telecontrol and tracing is reducing and the health consequences are reducing the concerns about privacy.

Figure 4: Relations among technologies and non-technologies tags

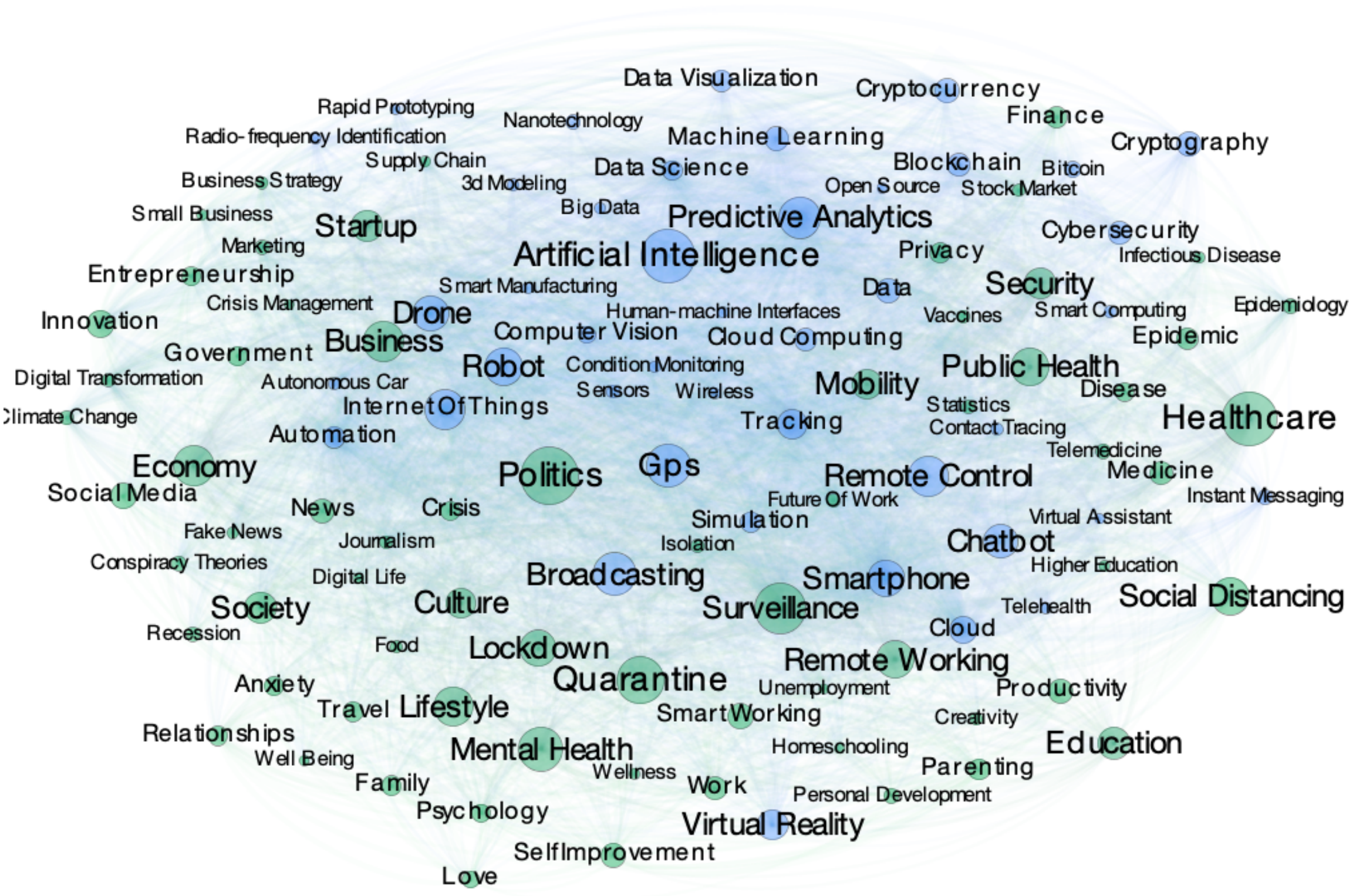

Figure 5: Graph of tags for articles concerning COVID-19 that contains a technology

The industrial side is not behaving differently. *Teleoperation* and *remote control* of production sites are becoming urgent. Owing to social distancing many companies (SMEs included) should organize their production in more shifts to maintain the same level of production they had before the COVID-19. However a plant organized to work with a certain number of workers and employees can not operate with half the number of personnel in two shift but a deep transformation is necessary. Machines, equipments, technical services etc.. should be substituted with modern ones (impossible due to the COVID-19 economical crisis) retrofitted or revamped. Industrial Internet of things seems to be the natural candidate for this transformation. A low cost approach based on *IoT devices* (at edge and node) and IT platforms (mainly small Manufacturing Execution Systems) is an affordable alternative to digital transformation.

In such a new scenario data coming from the machines will enable telecontrol of the factory, but also *remote* maintenance and *predictive maintenance*. Machine producers or plant installers as well as system integrators are struggling with teleoperating and acting on the machines and installations from remote. The goals are to detect what is happening at the plant/shop floor level and to teleoperate on the machine or guide the operator in performing some fixing, adjustment, assembly or spare part substitution. It opens the market to VR training for learning how to assembly and repair industrial machines to *headsets* with *AR and remote assistance*.

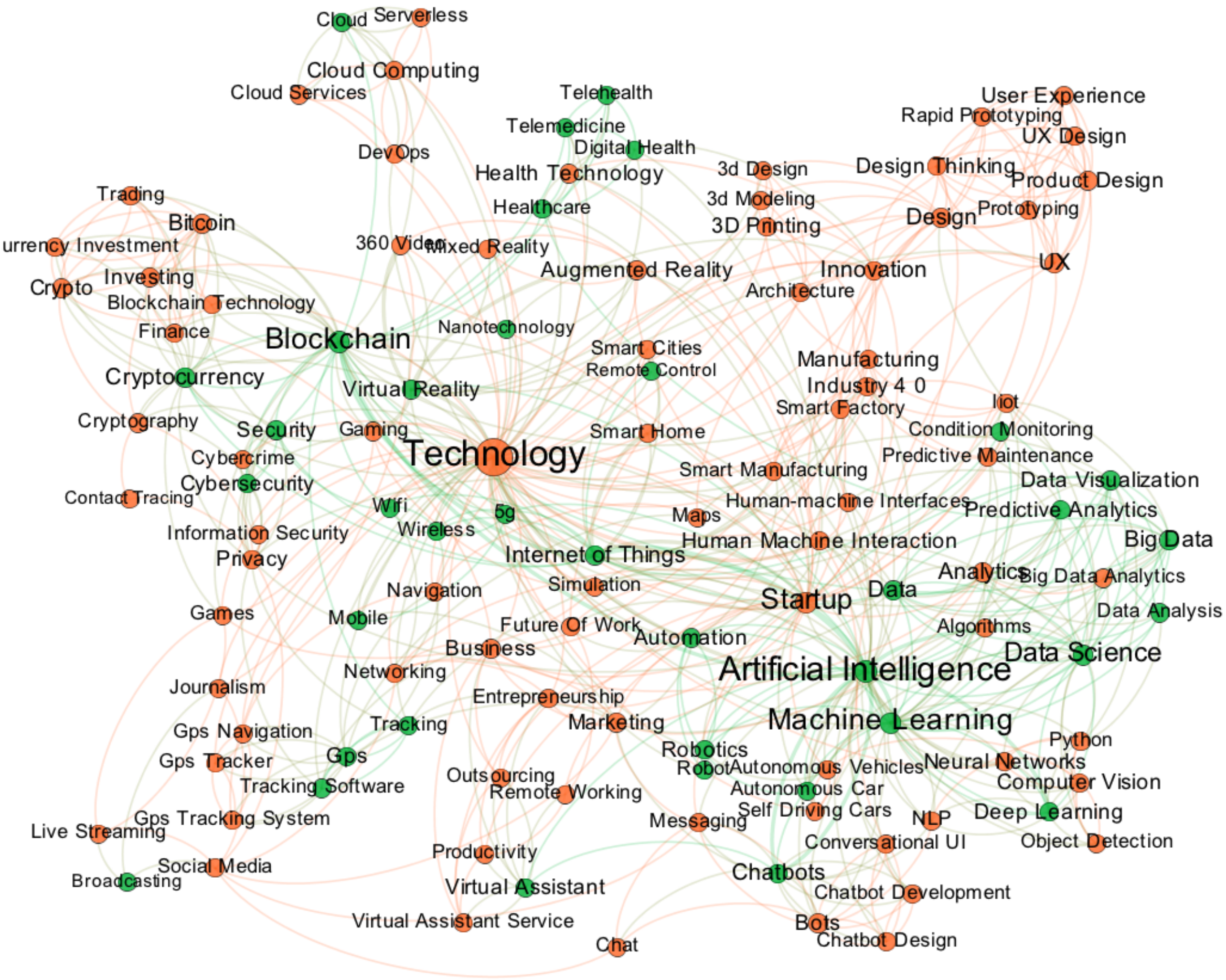

Figure 6: Graph of related tags of technologies

Robotic solutions are the other way to reduce social interactions and keep the production level. In particular in many SMEs *cobots* could play a role due to their simple programming and the reduced spaces. Moreover they are perceived mor as a tool or an assistant than as a robot, thus increasing the acceptance by employees.

Several regulations[3], sector standards[4] , company-syndicates' agreements (e.g. FCA with all the syndicates) impose a social distance also in working activities. It is oriented to reduce the virus transmission but really hard to achieve, control and use in case of infected people. Several devices appeared in the market: they are based on *bluetooth technology* on mobile phones or on small boards (embedded electronics) or on *RFID tags and antennas*.

[3]https://www.tracetogether.gov.sg
[4]https://www.lear.com/safeworkplaybookfiles

Singapore, with its TraceTogether[5] app, is applying people tracing and coronavirus diffusion based on mobile phone, QR code scanning, etc... The APP is for the general public and many other countries are rapidly following such a best practice but with limitation due to cultural and regulatory aspects (e.g. GDPR).

The case is different when a worker entres the company gate. No one can force the employees to install any app on their personal mobile phone. Moreover, smartphones are often forbidden in production to not disturb and distract the workers during their activities (sometimes dangerous). In other cases the use of cameras and smartphones are forbidden for IP reasons. therefore other alternatives have been proposed.

Several embedded solutions are spreading in industrial settings: from bracelet to keyrings, to pendants, to wearable wristbands. Many of them are conceived as alerting systems: from blinking leds to vibration. However all of them could perform more functions than the ones that now are accepted. Actually they can also record the contacts to trace the possible contacts with an infected person. Such a tracing activity could be perceived as a violation of GDPR and many national laws (e.g. the Italian law on workers' statute Article 4 of Law No 300 of 20/05/1970), which implemented the non traceability of workers' activities.

In perspective such devices can enable not only detection of events, but, with a more distributed architecture (several gateways for triangulation and localisation), also interesting analyses of assets movements into the shopfloor. Everything could be encrypted and tracking made anonymous. However the anonymous data (when analysed) can provide useful information, now for redefining facility layout for the COVID19 emergence, but later also for improving logistics and production.

The health emergence is pushing logistics to a new level (Amazon first demonstrated the acceleration). The explosion of online shopping exerted a huge pressure on the logistic chain that could now find further economies in asset tracing and tracking through beacon or passive/active RFID tags. *Blockchain* certified transactions and *smart contracts*, ready to emerge, could now find the occasion to overcome the friction in other fields than cryptocurrencies.

Last but not least, the so-called smart or remote working increased cybersecurity issues: fishing first and data breach later are the two most evident signs of unmanaged transitions. Actually even if collaboration tools and sharing platforms existed before and were adopted in companies, only few of them were really prepared to shift from working at the office to working from home. The COVID-19 demonstrated the low level of awareness for what concerns cybersecurity and the enormous quantum leap that companies have to make to address their cyber security aspects.

### 5.2 COVID, Technologies and their impact on Labor Market

Industry 4.0 was, and currently is, a strong and heterogeneous influence on the job market. A change in organisational activities and competences was a consequence to this phenomenon that clearly spreads differently worldwide at an exponential speed and with an uncertain impact [8]. Middle-sized and large enterprises saw the need for new assets and a change in the modus operandi that could be beneficial to growth and competitiveness. But what is and will be the impact of this industrial revolution on occupations and competences is still a sensitive topic open to discussions [9]; it is evident that, in the future, workers will be needed to be independent and to master operating new technologies, with additional cross-sectoral skills decisive for a market that seems to project towards digitisation.

However, the advent of this pandemic emergency brought a change to the way we deal with our daily life and work, and we cannot mistakenly isolate the two phenomena. First, the two phenomena are similar because of their impact on society. Second, the consequences of the pandemic are influencing the tech revolution. Within a few months, the pillars of Industry 4.0, (such as the role played by some new occupations, the need of digital skills required from enterprises or the fear of technology) are being redesigned.

The approach to work and the openness to changement will be shaped. First, if in the digital revolution the technology was seen as a potential cause of job destruction, in this situation it will be seen as support for the citizen since it is evident that it is an opportunity to mitigate the problems caused by forced distance in workplaces. Second, the borders between work time and leisure will change: the importance of managing stress due to hyperconnection, even outside the office will increase especially because of the physical boundaries in the work/non-work environment. Finally, this situation will lead people to transform their mentality towards accepting radical changes in the way they work; in particular, the fear of being replaced by the technology will give way to a greater open-mindedness.

A technology could be interpreted either as an enabler or a replacement for a working activity. In fact, a particular technology could help the worker to increase his productivity, but could also entirely replace his job. For these reasons, we have witnessed an increasing demand for adaptability with an emphasis on the concept of lifelong learning and human development [10]. With the advent of COVID-19, it is particularly necessary to have resilience and adaptability in order to face the new restrictions that the pandemic imposes. A new working reality leads to the consolidation of new skills and the rise of new job profiles. First, having people with digital skills is strictly necessary in order to adapt to smart working. Having digital skills in the company is no longer about competitive advantage, but about survival. Second, we will see the rise of profiles relating to risk management, ensuring compliance with safety regulations and specifically to biohazard monitoring. In particular, it will be the focus on soft skills such as time management, creativity and resilience.

[5] https://www.tracetogether.gov.sg/

However, we can state that the long debate on how much the introduction of technologies afflicts the labour market is definitely enriched by this renewed and exceptional dynamic: the advent of the COVID-19 sanitary emergency, can slow down or hasten the employment and spread of technologies. Even if this pandemic phenomenon is completely new, the effect on job market could be studied by looking on what was happening during the Industry 4.0 revolution.

However, we have to take into account the negative effects that the pandemic is having on the demand for labour. The shock of job demand will result in significant downward adjustments to wages and working hours. The lockdown is having a major impact on economic activities; by March 10, workers had lost about 30,000 hours of work, and for some of them this had resulted in lower wages. The negative impact on wages will also result in lower consumption of goods and services, which will necessarily also have an impact on activity turnover; poverty is therefore set to increase substantially [11]. In the face of this, while in Industry 4.0 it was possible to identify some professions more at risk, today entire social categories are vulnerable, for different reasons: young people not yet in employment, who will suffer the effects of the crisis; women, who are the most representative in the sectors most affected (such as services); unprotected workers, such as freelancers, who are much less protected than employees [12].

### 5.3 COVID, Technologies and their Social Impact

The COVID-19 crisis has forced the technological community to experience the first large scale exercise on creativity under constraints. We know from creativity studies and the neuroscience of creativity that meeting severe constraints may trigger novel solutions to a large degree. The crisis has forced technologists to work under negation of time (extreme time pressure), shortage of space (social distancing, close home environments), lack of materials (disruption of supply chains). Our data show that the ability of the technological community to react to the crisis by suggesting new solutions has been fast and effective.

Technologists working on Industry 4.0 solutions have found it, however, that the main functional requirements were, to a certain extent, similar at a high level of abstraction. Industry 4.0 solutions are based on the recombination between physical transformation activities (manufacturing) and intelligent (learning) information processing. These solutions habilitate the manipulation of physical reality at a distance, exploiting the power of telecommunication networks to transmit data at high speed.

The technologies needed to address the COVID-19 crisis require something similar, but in daily life. The physical contact must decrease to the minimum viable product in order to reduce or mitigate the risk of contagion. Consequently, the physical movement of people and goods must be reduced. The physical concentration of people in the same place must be avoided. In all these cases the remote operation of activities become crucial.

The analysis of technologies proposed under the crisis reveals a high level of conceptual flexibility in the community of inventors, startup companies, and technology-based firms. Having explored systematically the constraints of physical proximity in the remote operation of manufacturing activities has allowed technologists to develop rapidly adaptations and smart solutions. The social impact of technologies is the result of genuine inventivity and empathic engagement with large scale problems of humankind. This is good news.

## References

[1] K. Henning. Recommendations for implementing the strategic initiative industrie 4.0. 2013.

[2] Hussain H. I. Ślusarczyk B. Jermsittiparsert K. Haseeb, M. Industry 4.0: A solution towards technology challenges of sustainable business performance. *Social Sciences*, 2019.

[3] Xu E. L. Li L. Xu, L. D. Industry 4.0: state of the art and future trends. *International Journal of Production Research*, 2018.

[4] Trappey C. V. Govindarajan U. H. Sun J. J. Chuang A. C. Trappey, A. J. A review of technology standards and patent portfolios for enabling cyber-physical systems in advanced manufacturing. *IEEE Access*, 2016.

[5] Kreiner C. Macher G. Messnarz R. Riel, A. Integrated design for tackling safety and security challenges of smart products and digital manufacturing. *CIRP annals*, 2017.

[6] Trivelli L. Bonaccorsi A. Fantoni G. Chiarello, F. Extracting and mapping industry 4.0 technologies using wikipedia. *Computers in Industry*, 2018.

[7] Mathieu Bastian, Sebastien Heymann, and Mathieu Jacomy. Gephi: An open source software for exploring and manipulating networks. 2009.

[8] C. Last. Global commons in the global brain. *Technological Forecasting and Social Change*, 2017.

[9] Chiarello F. Coli E. Binda A. Fareri S., Fantoni G. Estimating industry 4.0 impact on job profiles and skills using text mining. *Computers in Industry*, 2020.

[10] Paul Schulte John Howard. The impact of technology on work and the workforce. *International Labour Organization*, 2019.

[11] International Labour Organization. Covid-19 and the world of work: Impact and policy responses. *ILO Monitor 1st Edition*, 2020.

[12] International Labour Organization. Covid-19 and the world of work. updated estimates and analysis. *ILO Monitor 2st Edition*, 2020.